\documentclass[aps,prb,twocolumn]{revtex4}
\usepackage{graphicx}
\usepackage{epsfig}
\usepackage{amsmath}
\usepackage{graphics}
\usepackage{epsfig}
\usepackage{tikz}
\usetikzlibrary{decorations.pathmorphing}

\usepackage{framed,color}
\definecolor{shadecolor}{rgb}{0.5, 1.0, 0.83}

\begin{document}

\title{Telegraph noise in Markovian master equation for electron transport through molecular junctions}
\author{Daniel S. Kosov}
\address{College of Science and Engineering,  James Cook University, Townsville, QLD, 4811, Australia }


\begin{abstract}
We present a theoretical approach to solve Markovian master equation for quantum transport with stochastic telegraph noise. 
Considering probabilities  as functionals of a random telegraph process we use the Novikov's functional method to convert stochastic master equation to a set of deterministic differential equations. The equations are then solved in the Laplace space and  the expression 
for the probability vector averaged over ensemble of realisations of stochastic process is obtained.
We apply the theory to study manifestations of  telegraph noise in transport properties of molecular junctions. We consider the quantum electron transport in a resonant-level molecule as well as polaronic regime transport in a molecular junction with electron-vibration interaction.
\end{abstract}

\maketitle

\section{Introduction}

Recently, there have been a significant experimental and theoretical interest in studying  the fluctuations  of the electric properties  of molecular junctions.\cite{avriller09,thoss14,segal15,PhysRevLett.100.196804,rudge16a,kosov17-wtd,kosov17-nonren,doi:10.1021/nl201327c,doi:10.1021/nl060116e,Tsutsui:2010aa,diventra05,galperin06a,PhysRevLett.101.046801,belzig09,brandbyge12,PhysRevB.87.115407}
There are two distinctly different types of stochastic fluctuations in current-carrying molecular junctions. On the fundamental level,
there is always noise due to the probabilistic nature of quantum mechanics.\cite{nazarov-book}  Moreover, 
the electron transport  in molecular junctions often demonstrate random switching of the electric current between multiple different values 
due to  presence of  localised electronic states,  various  geometric
stochastic conformational switching, and current-triggered irregular motion of  interface atoms.\cite{noise10,noise14,telegraph16,noise17,nichols2010}

The scope of theoretical work  on telegraph noise in steady state quantum transport is still very limited.\cite{telegraph94,telegraph95,gurvitz16,telegraph17} The resonant electron  tunneling through  a "two-level fluctuator" was studied more than 20 years ago using the Green's function methods.\cite{telegraph94,telegraph95} Then, only recently, Gurvitz et al. described both time-dependent and steady state electric current through the single-electron transistor with  random Markovian jumps of the single-electron energy level.\cite{gurvitz16} The approach, which incorporates telegraph noise into Keldysh nonequilibrium Green's functions, was proposed last year to study quantum heat transport through  a single fluctuating level electronic junction.\cite{telegraph17}
In this paper we develop a statistical theory of electron transport in molecular junctions with time-varying parameters,  which can switch stochastically between two different values (telegraphic noise).   We work with the Markovian master equation
and we incorporate  the telegraph process into electron transfer rates - it results into the stochastic master equation. The stochastic differential equations, which represent the master equation with telegraph noise,  are converted to the closed system of deterministic differential equations by using Novikov's method of stochastic functionals.\cite{novikov64} The method was first proposed by Novikov in 1964 in theory of
of turbulence\cite{novikov64} and then developed in details by Klyatskin in the series of monographs.\cite{klyatskin2005,klyatskin2005a,klyatskin2011} 
The similar approach was recently employed to study electron transport through single fluctuating energy level.\cite{gurvitz16}
Telegraph  stochastic processes considered in this paper  are closely related to Markovian kinetic equations with  Keilson-Storer kernel,\cite{KS1954} which have many application in chemical kinetics.\cite{Burshtein,gelin96,Gelin:1999aa,gelin:144514,GELIN201535,shushin}

The paper is organised as follows. Section II describes the solution of the general Markovian master equation with rates, which depend on a stochastic telegraph process.  In Section III, we apply the theory to the problem of electron transport though a molecular junction: first, we consider a simple case of single resonant molecular orbital and, second, we discuss the telegraph noise in a molecular junction with electron-vibrational interaction.
Section IV summarises the main results of the paper.

We use natural  units for quantum transport  throughout the paper: $\hbar=k_{B}=e=1$.

\section{Markovian master equation with telegraph noise}
In this section we consider a general problem of  finding the solution for master equation with time dependent rates which undergo jumps between two different values at specific but ransom times (telegraph noise). 
We discuss general theory first (applicable to any Markovian dynamics with telegraph noise) and then apply the theory to quantum electrons transport through molecular junction in the next section.
We begin with the Markovian master equation
\begin{equation}
P_k(t) =\sum_{k'} W_{kk'} P_{k'}(t),
\label{me0}
\end{equation}
where $P_k(t)$ is the probability of observing our system in state $k$ at time $t$ and $W_{kk'}$ is the transition probability per unit time (rate) from state $k'$ to state $k$.
In matrix form master equation (\ref{me0})  can be compactly written 
\begin{equation}
 \frac{d \mathbf P(t)}{dt}  = \mathbf W  \mathbf P(t).
 \label{me1}
\end{equation}

Suppose that the parameters of the system experience some stochastic variations,  which result into the telegraph noise in the rates of the master equation. The master equation becomes
\begin{equation}
 \frac{d \mathbf P(t)}{dt}  = \mathbf W  \mathbf P(t)  + \xi(t)  \mathbf A \mathbf P(t).
 \label{me2}
\end{equation}
 Here  function $\xi(t)$ describes  the stochastic telegraph noise process,  the function switches stochastically between  two values as time progresses
\begin{equation}
\xi(t)= \gamma (-1)^{n(0,t)},
\end{equation}
where $n(t,t')$ is the random sequence of integer numbers describing the number of telegraph jumps within the time interval $[t,t']$.\cite{klyatskin2011}
It means that   dynamics of the  probabilities is governed by the time-dependent matrix that switches stochastically between 
$  \mathbf W + \mathbf A $ and $ \mathbf W - \mathbf A $, and the probability vector itself $\mathbf P(t)$ becomes a stochastic function of time.  The matrix $\mathbf A$ is not  completely arbitrary, it should have two general properties of $\mathbf W$-matrices of a Markovian master equation\cite{vanKampen}
\begin{equation}
W_{kk'}\pm \gamma A_{kk'} \ge 0 \text{  for   } k\ne k'
\label{positivity}
\end{equation}
to preserve  positivity of the probabilities
and 
\begin{equation}
\sum_k A_{kk'} = 0 \text{  for any   }  k'
\end{equation}
to maintain  proper normalization of the probability vector at each moment of time despite of the stochastic discontinuous changes of the rate matrix.

We assume that the distribution of times at which telegraph jumps to occur is the Poisson point distribution. It means that 
the probability to observe $n$ jumps during the time interval $[t,t']$ is
\begin{equation}
p_{n(t,t')=n} = \frac{(\alpha (t-t'))^n}{n!} e^{-\alpha (t-t')}
\end{equation}
with average number of telegraph jumps during the time interval
\begin{equation}
\overline{n(t,t')} =  \sum_{n=0}^{\infty} n p_{n(t,t')=n}  = \alpha (t-t').
\end{equation}
Using the Poisson point distribution, we compute the first few moments of the stochastic function $\xi(t)$ averaging over the realisations of the stochastic process.
The average value of the stochastic process $\xi(t)$ at time t is
\begin{equation}
\big\langle \xi(t) \big\rangle_{\xi}  = \sum_{n=0}^\infty p_{n(0,t)=n} \gamma (-1)^{n}= \gamma e^{-2 \alpha t},
\label{m1}
\end{equation}
where $\big\langle ... \big\rangle_{\xi} $ means the averaging over all possible realization of stochastic process $\xi(t)$.
Two-time  correlation function for  $t_1>t_2$ is 
\begin{multline}
\big\langle \xi(t_1)  \xi(t_2)\big\rangle_{\xi}  =  \big\langle \gamma (-1)^{n(0,t_1)} \gamma (-1)^{n(0,t_2)} \big\rangle_{\xi} \\
= \gamma^2 \big\langle (-1)^{n(t_2,t_1)}   \big\rangle_{\xi} = \gamma^2 e^{-2 \alpha (t_1-t_2)},
\label{m2}
\end{multline}
and all higher order correlation functions can be factorised
in terms of average process and two-time correlation functions.\cite{klyatskin2011}

Let us now 
average the stochastic master equation (\ref{me2}) over the ensemble of telegraphic noise realisations
\begin{equation}
 \frac{d}{dt} \big\langle  \mathbf P(t) \big\rangle_{\xi}  = \mathbf W  \big\langle  \mathbf P(t)  \big\rangle_{\xi}  +  \mathbf A  \big\langle \xi(t)  \mathbf P(t) \big\rangle_{\xi}.
 \label{equation1}
\end{equation}
Differential equation (\ref{equation1})  is not yet in a  complete form since
 we do not know  how to compute $\langle \xi(t)  \mathbf P(t) \rangle_{\xi}$.

To evaluate $ \mathbf \langle \xi(t) \mathbf P(t) \rangle_{\xi}$ 
we notice that  $\mathbf P(t)$ can be considered as a functional of the stochastic process $\xi(t)$ 
\begin{equation}
\mathbf P(t) = \mathbf P[\xi](t).
\end{equation}
Next we use
Novikov's method of computing averages of products of  functional of stochastic process with the process function itself. \cite{novikov64,klyatskin2005,klyatskin2005a,klyatskin2011}
Following Novikov ideas\cite{novikov64} we introduce
\begin{equation}
\Big \langle \xi(t)  \mathbf P[\xi+z](t) \Big\rangle_{\xi},
\end{equation}
where $z(t)$ is an arbitrary  reasonably smooth deterministic function.
We expand $\mathbf P[\xi +z](t)$ in the Taylor-like series using functional differentiation around $\xi(t)=0$
\begin{equation}
\mathbf P[\xi+z](t) = \exp \Big\{ \int_{0}^{\infty} dt_1 \xi(t_1) \frac{\delta}{\delta z(t_1)} \Big\} \; \mathbf P[z](t).
\label{pz}
\end{equation}
We notice that $\mathbf P[z](t)$ is a completely deterministic functional and all stochastic variables are now localised in the operator exponent  in (\ref{pz}), therefore
\begin{multline}
 \Big\langle \xi(t)  \mathbf P[\xi+z](t)  \Big\rangle_{\xi} \\
 =  \Big\langle \xi(t)  \exp \Big\{ \int_{0}^{\infty} dt_1 \xi(t_1) \frac{\delta}{\delta z(t_1)} \Big\}  \Big\rangle_{\xi} \;  \mathbf P[z](t).
 \label{eqa0}
\end{multline}
\begin{widetext}
Expanding the operator exponent in (\ref{eqa0}), we get
\begin{multline}
\Big\langle \xi(t)  \mathbf P[\xi+z](t) \Big\rangle_{\xi}=  \Big\langle \xi(t)  \Big[1+ \sum_{k=1}^\infty \frac{1}{k!} \int_{0}^{\infty} dt_1 \int_{0}^{\infty} dt_2 .... \int_{0}^{\infty} dt_k \xi(t_1) \xi(t_2) ... \xi(t_k) 
\frac{\delta}{\delta z(t_1)} \frac{\delta}{\delta z(t_2)} ... \frac{\delta}{\delta z(t_k)} \Big] \Big\rangle_{\xi} \;  \mathbf P[z](t) 
\\
=   \Big[\Big\langle \xi(t) \Big\rangle_{\xi} +  \sum_{k=1}^\infty  \int_{0}^{\infty} dt_1 \int_{0}^{t_1} dt_2 .... \int_{0}^{t_{k-1}} dt_k    \Big\langle \xi(t) \xi(t_1) \Big\rangle_{\xi}   \Big\langle  \xi(t_2) ... \xi(t_k)  \Big\rangle_{\xi} 
\frac{\delta}{\delta z(t_1)}\frac{\delta}{\delta z(t_2)} ... \frac{\delta}{\delta z(t_k)} \Big]  \;  \mathbf P[z](t) 
\\
=   \Big[ \Big\langle \xi(t) \Big\rangle_{\xi} +    \int_{0}^{\infty} dt_1    \Big\langle \xi(t) \xi(t_1) \Big\rangle_{\xi} \frac{\delta}{\delta z(t_1)}
\Big\langle \exp \Big\{ \int_{0}^{t_1} dt' \xi(t') \frac{\delta}{\delta z(t')} \Big\} \Big\rangle_{\xi} \Big]  \;  \mathbf P[z](t).
\end{multline}
Substituting expressions for the moments  of the stochastic process (\ref{m1},\ref{m2}) yields
\begin{equation}
\langle \xi(t)  \mathbf P[\xi+z](t) \rangle_{\xi}
= \Big[  \gamma e^{-2 \alpha t} +   \gamma^2 \int_{0}^{\infty} dt_1    e^{-2 \alpha (t-t_1)} \frac{\delta}{\delta z(t_1)}
\Big\langle \exp \Big\{ \int_{0}^{t_1} dt' \xi(t') \frac{\delta}{\delta z(t')} \Big\} \Big\rangle_{\xi} \Big]  \;  \mathbf P[z](t).
\end{equation}
We differentiate the lhs and the rhs of this equation with respect to $t$ and obtain
\begin{equation}
(\frac{d}{dt} + 2 \alpha) \langle \xi(t)  \mathbf P[\xi+z](t) \rangle_{\xi} = \Big[  \gamma e^{-2 \alpha t} +   \gamma^2 \int_{0}^{\infty} dt_1    e^{-2 \alpha (t-t_1)} \frac{\delta}{\delta z(t_1)}
\Big\langle \exp \Big\{ \int_{0}^{t_1} dt' \xi(t') \frac{\delta}{\delta z(t')} \Big\} \Big\rangle_{\xi} \Big]   \; \frac{d}{dt} \mathbf P[z](t),
\end{equation}
\end{widetext}
which is simply
\begin{equation}
(\frac{d}{dt} + 2 \alpha) \langle \xi(t)  \mathbf P[\xi+z](t) \rangle_{\xi} =    \langle \xi(t)  \;\frac{d}{dt} \mathbf P[\xi+z](t) \rangle_{\xi}.
\end{equation}
Setting $z(t)=0$ in the above equation, we get\cite{shapiro-loginov,klyatskin2011}
\begin{equation}
(\frac{d}{dt} + 2 \alpha) \langle \xi(t)  \mathbf P[\xi](t) \rangle_{\xi} =    \langle \xi(t)  \;\frac{d}{dt} \mathbf P[\xi](t) \rangle_{\xi}.
\label{eqa}
\end{equation}
Eq.(\ref{eqa}) is sometimes called the  Shapiro-Loginov differential formula.\cite{shapiro-loginov,gurvitz16} 
Substituting $\frac{d}{dt} \mathbf P[\xi](t)$ from the initial master equation (\ref{me2}) into right hand side of (\ref{eqa})
gives
\begin{equation}
(\frac{d}{dt} + 2 \alpha-\mathbf W) \langle \xi(t)  \mathbf P(t) \rangle_{\xi} =  \gamma^2 \mathbf A  \langle \mathbf P(t) \rangle_{\xi}.
\label{equation2}
\end{equation}
We omit  here  and throughout the rest of the paper  the functional dependence notation in $\mathbf P$ for brevity. 
Note that although (\ref{eqa}) can be generalized to an arbitrary Gauss-Markov stochastic process,\cite{shapiro-loginov} equation
 (\ref{equation2})  already requires that  $\xi(t)$ should be  a dichotomic process in order to close the hierarchy of the differential equations at $\langle \xi  \mathbf P \rangle_{\xi}$ term.

 We have converted the initial stochastic master equation (\ref{me2}) to the closed system of two coupled deterministic differential equations (\ref{equation1},\ref{equation2}).
Next we introduce function
\begin{equation}
\mathbf F(t)= \langle \xi(t) \mathbf P(t) \rangle_{\xi}
\end{equation}
and perform the Laplace transformation on the system of differential equations (\ref{equation1},\ref{equation2}). We find
\begin{equation}
s \langle \widetilde \mathbf P(s) \rangle_{\xi} - \mathbf P(0)  = \mathbf W  \langle  \widetilde \mathbf P(s)  \rangle_{\xi} +  \mathbf A \widetilde \mathbf F(s),
\label{equation1L}
\end{equation}
\begin{equation}
(s \mathbf I + 2 \alpha-\mathbf W) \widetilde \mathbf F(s)  - \gamma \mathbf P(0)=  \gamma^2 \mathbf A  \langle \widetilde \mathbf P(s) \rangle_{\xi},
\label{equation2L}
\end{equation}
where "tilde"  denotes functions in the Laplace space, for example, $ \widetilde \mathbf P(s)= \int_0^\infty dt e^{-st}  \mathbf P(t)$ and $\mathbf I$ is the identity matrix.
While deriving (\ref{equation1L},\ref{equation2L}) we took into account that the probability vector $\mathbf P(t)$ does not depend on stochastic varaible $\xi$  at initial time $t=0$,  therefore   $\langle  \mathbf P(0) \rangle_{\xi}  =   \mathbf P(0)$ and 
\begin{equation}
\langle  \mathbf \xi(0) \mathbf P(0) \rangle_{\xi}  =  \langle  \mathbf \xi(0)  \rangle_{\xi} \mathbf P(0)  =\gamma \mathbf P(0).
\end{equation}
Eliminating  $\widetilde \mathbf F(s)$ 
from (\ref{equation1L},\ref{equation2L}) gives the following expression for the noise averaged  probability vector in Laplace space
\begin{widetext}
\begin{equation}
\label{ps}
\langle \widetilde \mathbf P(s) \rangle_{\xi} = \left [ (s \mathbf I -\mathbf W) -\gamma^2 \mathbf A \frac{1}{ (s +2 \alpha) \mathbf I -\mathbf W }\mathbf A \right]^{-1}\Big[ \mathbf I + \gamma \mathbf A \frac{1}{ (s +2 \alpha) \mathbf I -\mathbf W }\Big]  \mathbf P(0).
\end{equation}
\end{widetext}

\section{Applications to the quantum  transport}
\subsection{Electron transport through single-resonant level with telegraph noise in the contacts }
Let us consider the electron transport through a molecular junction represented by a single resonant molecular orbital connected to two leads (left and right) held at different chemical potentials.
The electron spin is not considered, therefore the molecular orbital can only contain zero or one electron.
The Markovian master equation is given by (\ref{me2}),
where matrix $\mathbf W$   is
\begin{eqnarray}
\mathbf W= \left[
 \begin{array}{cc}
   -  T_{10} &  T_{01}  \\  
     T_{10}  & - T_{01} 
    \end{array} 
    \right],
    \label{w}
\end{eqnarray}
and $ \mathbf P(t) $ is the probability vector  
\begin{eqnarray}
\mathbf P(t)= \left[
 \begin{array}{c}
   P_0(t) \\  
   P_1(t)
    \end{array} 
    \right].
\end{eqnarray}
Here  $P_0(t)$ is the probability that the molecular orbital has zero electrons at time $t$ and  $P_1(t)$ is the probability  for the orbital to be occupied by one electron at time $t$.  The dynamical evolution of these probabilities is determined by two rates: $T_{01}$ describes the transition from occupied to empty state and  $T_{10}$ is the rate for the opposite process.
The total rates consist of two  contributions from  left and right electrodes:
\begin{equation}
 T_{10} =  T^L_{10} +  T^R_{10},  \;\;\;\;  T_{01} =  T^L_{01} +  T^R_{01}.
\end{equation}
The partial rates are given by the standard Fermi golden rule expressions
\begin{eqnarray}
 T^L_{01} &=&  \Gamma_L f_L,\;\;\;\;  T^R_{01} =  \Gamma_R f_R,
\\
 T^L_{10} &= &  \Gamma_L (1- f_L), \;\;\;\;  T^R_{10} =  \Gamma_R (1-f_R),
\end{eqnarray}
where $ \Gamma_{L/R} $ are molecular level energy broadening functions due to coupling to  left/right electrode. 
Fermi occupation numbers for left and right electrodes are
\begin{equation}
f_L =[1+e^{(\epsilon-\mu_L)/T_L}]^{-1}, \;\;\;\; f_R= [1+e^{(\epsilon-\mu_R)/T_R}]^{-1}.
\end{equation}
Voltage bias is defined as the difference between  left and right electrodes  chemical potentials:
$V=\mu_L -\mu_R$.

Let us introduce the telegraph noise into the master equation. Recently Gurvitz et al.\cite{gurvitz16} used similar method to model a single electron transistor with energy level which can stochastically jump between two given values.
The most prevailing scenario for molecular junctions is not the fluctuation of the molecular orbital itself but rather the telegraph noise  originating from various processes on the molecule-metal interfaces.\cite{noise10}
To model this setting we assume
 that the coupling between the molecule and the left electrode $ \Gamma_L $  undergoes telegraph type stochastic oscillations (the choice of left electrode is arbitrary, we can also consider the  coupling between molecule and the right electrode $\Gamma_R$).
\begin{equation}
\Gamma_L(t) = \Gamma_L (1+ \xi(t))
\end{equation} 
Here $\xi(t)$ is a stochastic telegraph noise variable, which can switch between two states. Therefore, matrix $\mathbf A$ is
 \begin{eqnarray}
\mathbf A= \left[
 \begin{array}{cc}
   -  T^L_{10} &  T^L_{01}  \\  
     T^L_{10}  & - T^L_{01} 
    \end{array} 
    \right].
    \label{a}
\end{eqnarray}
With this definition of matrix $\mathbf A$ the amplitude of the stochastic telegraph process $|\gamma|$ should be $\le1$ to preserve the positivity of the probabilities (\ref{positivity}).

The limit $\lim_{s\rightarrow0} s \widetilde f(s)$ of  the Laplace transformed function corresponds to the asymptotic time limit of the real time function   
$\lim_{t\rightarrow\infty}  f(t)$. 
Therefore, 
the stationary (nonequilibrium steady state) probability vector is 
\begin{equation}
\langle \mathbf P \rangle_\xi = \lim_{s\rightarrow0} s  \langle \widetilde \mathbf P(s) \rangle_{\xi} 
\end{equation}
Substituting matrices $\mathbf W$ (\ref{w}) and $\mathbf A$ (\ref{a}) into the expression for the probability vector in the Laplace space (\ref{ps}), then
multiplying it by $s$ and letting $s$ tend to zero, we obtain the
 steady state probability vector
\begin{widetext}
\begin{eqnarray}
\langle \mathbf P \rangle_\xi =
   \frac{1}{4\Gamma
   (\Gamma+ \alpha ) - \Gamma_L^2 \gamma ^2}
\left[
\begin{array}{c}
2(\Gamma+ \alpha ) \{\Gamma_L
   (1-f_L)+\Gamma_R (1-f_R)\} - \Gamma_L^2 \gamma ^2 (1-f_L) 
   \\
 2(\Gamma+ \alpha )\{ \Gamma_L f_L+\Gamma_R f_R \} -\Gamma_L^2 \gamma ^2 f_L
   \end{array}
   \right],
\end{eqnarray}
\end{widetext}
where we introduced  $\Gamma=(\Gamma_L + \Gamma_R)/2$.
Notice that any dependence on initial probability vector $\mathbf P(0)$ completely disappear from the expression in the steady state.
In the noiseless limit ($\gamma=0$), we get
\begin{eqnarray}
\langle \mathbf P \rangle_\xi =
   \frac{1}{2\Gamma}
\left[
\begin{array}{c}
\Gamma_L
   (1-f_L)+\Gamma_R (1-f_R)
   \\
 \Gamma_L f_L+\Gamma_R f_R 
   \end{array}
   \right],
\end{eqnarray}
which is the correct standard expression for the occupation probabilities of a single resonant-level.

The current is computed with the use of the continuity equation. The average number of electrons in the system is
\begin{equation}
\langle N(t) \rangle_\xi =\langle P_1(t)\rangle_\xi.
\end{equation}
The continuity equation
\begin{equation}
\frac{d }{dt}\langle P_1(t)\rangle_\xi = J_L(t) + J_R(t)
\end{equation}
is used to identify expressions for the electron current
\begin{equation}
\langle J \rangle_\xi=\langle J_L \rangle_\xi= - \langle J_R \rangle_\xi= \Gamma_R [-f_R,   (1-f_R)]\cdot \langle \mathbf P \rangle_\xi.
\label{j1}
\end{equation}
Computing the dot product in (\ref{j1}), we get
\begin{equation}
\langle J \rangle_\xi = \frac{\Gamma_L \Gamma_R  (2 \Gamma+2 \alpha  -\Gamma_L \gamma^2 )}{4 \Gamma (\Gamma+ \alpha ) -\Gamma_L^2 \gamma^2} (f_L -f_R).
\end{equation}

\begin{figure}
\begin{center}
\includegraphics[width=1\columnwidth]{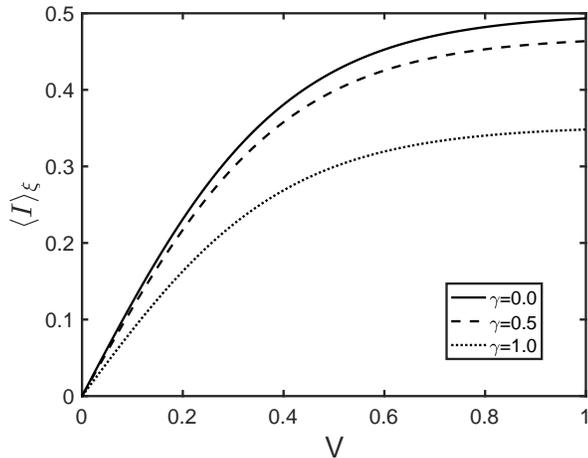}
\end{center}
\caption{Noise averaged current  for single resonant-level as a function of the applied voltage bias for different values of the amplitude of the telegraph noise $\gamma$.
Parameters used in calculations: $\Gamma_R =\Gamma_L = \Gamma=1$, $T=0.1$, and $\epsilon=0$ - all energy units are given in terms of $\Gamma$.  Unit for for electric current is $\Gamma$ (or if we put $\hbar$ and $e$ back, it is  $ e \Gamma/\hbar$) and  values of voltage bias $V_{sd}$ are given in $\Gamma$. }
\label{figure1}
\end{figure}

\begin{figure}
\begin{center}
\includegraphics[width=1\columnwidth]{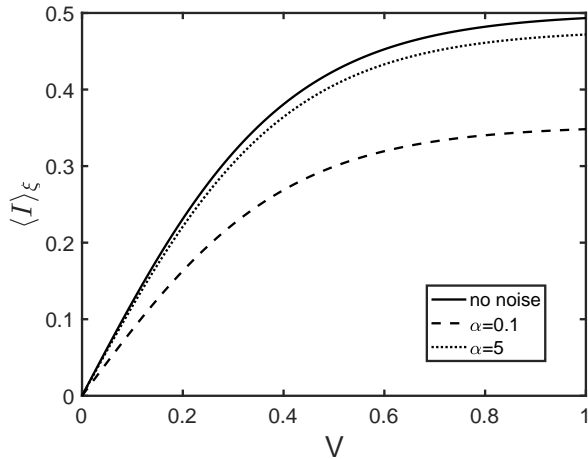}
\end{center}
\caption{Noise averaged current for single resonant-level as a function of the applied voltage bias for different values of the number of telegraph jumps per unit time 
$\alpha$. Parameters and units are the same as in figure 1.}
\label{figure2}
\end{figure}

Figure 1  and 2 show the current-voltage characteristics of a single resonant-level averaged over realisations of the stochastic telegraph process.
The noise in molecular electrode contact reduces the electric current and also reduces the molecular conductivity (the slope of I-V curve).  The reduction of the current is more significant  for  the larger amplitudes of the telegraph noise $\gamma$. The rare telegraph processes (the small density of  telegraph jump points on the time axis, $\alpha$) give the larger  reduction and very frequent jumps bring the current to the noiseless limit.To get physical insight we may consider the ratio  $\alpha/\Gamma$ as the number of telegraph jumps per average time required for an electron to tunnel across the molecule. Therefore, with our choice of the parameters, $ \alpha=5$ roughly corresponds to 5 telegraph switching per one tunnelling electron and $\alpha=0.1$ represents one telegraph jump during approximately 10 electron tunnelling events.

\subsection{Telegraph noise in molecular junctions with electron-vibration coupling}

Let us consider the case of the  electron transport through a molecular junction with electron-vibration interaction. The molecular bridge itself is described by Holstein Hamiltonian
\begin{equation}
H_{\text{molecule}}= \epsilon_0 a^\dag a  + \lambda \omega (b^\dag + b) a^\dag a +  \omega b^\dag b,
\end{equation}
where $\epsilon_0$ is molecular orbital energy,  $\omega$ is molecular vibration energy, and $\lambda$ is the strength of the electron-vibration coupling. $a^\dagger (a) $  creates (annihilates) an electron on the molecular orbital, and $b^+ (b)$ is bosonic creation (annihilation) operator for the molecular vibration. 

We assume  that the vibration maintains the equilibrium Bose-Einstein distribution at all time and, as a result, we obtain the  thermally averaged rates for the electron transfers ($m,n=0,1$)
\begin{equation}
 T^{L/R}_{mn} = \sum_{qq'} \Gamma^{L/R}_{mq,nq'} \frac{e^{-q' \omega/T}}{1-e^{-\omega/T}}.
 \label{r}
\end{equation}
The individual  transition rates between microscopic molecular states are computed using Fermi golden rule.\cite{PhysRevB.69.245302}
\begin{equation}
\Gamma^{L/R}_{0q',1q} =  \Gamma_{L/R} |X_{q'q}|^2 \left(1-f_{L/R}[\epsilon-\omega (q'-q)] \right)
\end{equation}
is the rate for the  transition from state occupied by one electron and $q$ vibrations to the electronically unoccupied state with $q'$ vibrations  by the electron transfer from the molecule to left and right  electrodes, respectively,
and
\begin{equation} 
\Gamma^{L/R}_{1q',0q} =    \Gamma_{L/R} |X_{q'q}|^2  f_{L/R}[\epsilon+\omega (q'-q)]
\end{equation}
is the rate for the  transition when electron is transferred from left/right  electrode into the originally empty molecules simultaneously changing the vibrational state from $q$ to $q'$. These rates depend on 
the electronic level broadening functions
$ \Gamma^{L/R}$, the Fermi occupation numbers $f_{L/R}$,
and  
the Franck-Condon factors
$X_{qq'}$.

The
matrices $\mathbf  W$  and  $\mathbf A$ have the same structure as in the transport through a single resonant-level, that are  (\ref{w})  and  (\ref{a}) with the transitional rates defined by (\ref{r}). Likewise to the electron transport through a single resonant-level, we compute steady state probabilities and electric current which are averaged over the realisations of the telegraph process.

\begin{figure}
\begin{center}
\includegraphics[width=1\columnwidth]{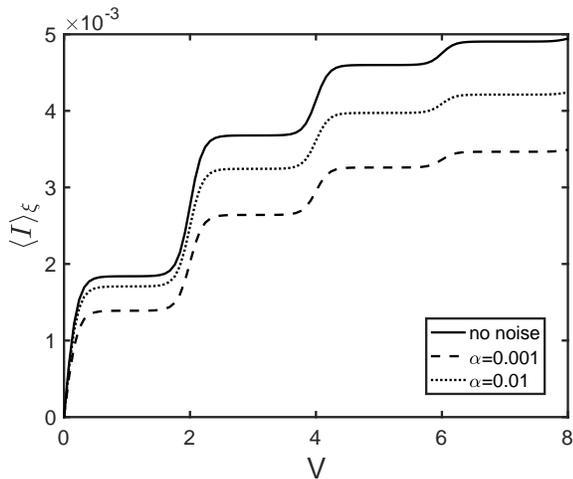}
\end{center}
\caption{Noise averaged current in a molecular junction with electron-vibration interaction as a function of the applied voltage bias computed  for different values of the number of telegraph jumps per unit time. Parameters used in the calculations: $\Gamma_R = 0.01$,
$\Gamma_L = 0.01$, $T=0.05$, $\lambda=1$, $\omega=1$, $\epsilon=0$ -- all energy units are measured in terms of vibrational frequency energy $\omega$.  Unit for for electric current is $\omega$ 
(or if we put $\hbar$ and $e$ back, it is  $ e \omega/\hbar$) and  values of voltage bias $V_{sd}$ are given in $\omega$.}
\label{figure3}
\end{figure}

Figure 3 shows the current as a function of the applied voltage computed for different densities of telegraph jump points on the time axis. The steps in the current correspond to the resonant excitations of the vibration states by electric current which occur when the voltage passes through an integer multiple of the vibration energy.\cite{PhysRevB.83.115414,fcblockade05,PhysRevB.69.245302}  As in the single resonant-level case considered before, the telegraph noise in  the molecule-electrode contacts results in the reduction of the electric current, but the frequent switching events (large $\alpha$) again brings the I-V curve to the optimal noiseless limit.

\section{Conclusions}

We have developed a theoretical approach to solve Markovian master equations with a telegraph noise. The rates of the master equation depend on a set of time-dependent parameters, which switch randomly between two values. We treated the probabilities as functionals of a stochastic telegraph process. We used the Novikov's method, originally developed in the theory of turbulence, to shift the stochastic functionals to  deterministic function domain; this procedure converted the stochastic master equation to a closed system of deterministic differential equations. The equations were solved with the use of the Laplace transformation and the general expression for the probability vector averaged over realisations of the stochastic process was derived. The theory was applied to quantum electron transport through model molecular junctions. We studied the role of the noise in the molecule-electrode contacts on the current-voltage characteristics.  Two models were considered: a single resonant-level transport and the Holstein model in the polaronic regime. In both models, the contact telegraph noise reduces the electric current and lowers the molecular conductivity.



\end{document}